\documentclass[aps,12pt,floatfix,final,notitlepage,oneside,onecolumn,nobibnotes,
nofootinbib,superscriptaddress,noshowpacs,centertags]{revtex4}

\RequirePackage{amssymb,amsmath} \RequirePackage{eufrak}
\RequirePackage{mathtext} \RequirePackage[cp1251]{inputenc}
\RequirePackage[T2A]{fontenc}
\RequirePackage[russian,english]{babel} \RequirePackage{bm}
\RequirePackage{array} \RequirePackage{longtable}
\RequirePackage{graphicx} \RequirePackage{calc}
\RequirePackage{ifthen} \RequirePackage{caption2}
\RequirePackage{subfigure}

\def\refitem#1{\relax}

\begin{document}

\selectlanguage{english}

\title{Radial HI Profiles at the Periphery of Galactic Disks:
The Role of Ionizing Background Radiation}

\author{\firstname{~O.~V.} \surname{Abramova}}
\email[]{foxirmos@gmail.com} \affiliation{Moscow State University, Sternberg Astronomical Institute, Universitetskii pr. 13, Moscow, 119992 Russia}

\begin{abstract}
Observations of neutral hydrogen in spiral galaxies reveal a sharp cutoff in the radial density
profile at some distance from the center. Using 22 galaxies with known HI distributions as an example, we
discuss the question of whether this effect can be associated exclusively with external ionizing radiation, as
is commonly assumed. We show that before the surface density reaches $\sigma_{\textrm{HI}}\le 0.5~{\cal M}_\odot/{\textrm {pc}}^2$(the same for galaxies of different types), it is hard to expect the gas to be fully ionized by background radiation. For two
of 13 galaxies with a sharp drop in the HI profile, the “steepening” can actually be caused by ionization. At
the same time, for the remaining galaxies, the observed cutoff in the radial HI profile is closer to the center
than if it was a consequence of ionization by background radiation and, therefore, it should be caused by
other factors.\\
{\bf Article is published in {\it Astronomy Letters, 2012, Volume 38, Number 4, Pages 222-230}.}
\end{abstract}

\maketitle

\section{INTRODUCTION}

\hspace{0.6cm}Neutral hydrogen in galaxies extends to large distances from the center exceeding considerably the optical radius $R_{25}$. At the same time, the surface density of neutral hydrogen in spiral galaxies drops sharply at some distance from the center. The ionization of a neutral gas by ultraviolet background radiation was considered as an explanation of this phenomenon (see Bochkarev and Sunyaev 1977; Maloney 1993; Corbelli and Salpeter 1993). However, the number of galaxies considered in the cited papers is small; in addition, fairly rough estimates of the volume densities for the gaseous and stellar disks were used in them. The elimination of these shortcomings makes this work topical.

At the disk periphery, where there is virtually no star formation, the contribution of young stars to ionizing radiation is nearly zero, and the contribution
of background radiation is decisive. According to Haardt and Madau (1996), quasars at high redshifts are mainly the sources of ionizing background radiation.
Direct observations of background radiation are difficult to perform, because it is absorbed by a thick layer of atomic hydrogen in the Galactic disk.

Starting from some distance from the center, the bulk of the hydrogen must be ionized. In this paper, we make an attempt to reconstruct the “total” (neutral+ionized) hydrogen profiles to estimate the total surface density in the outer regions of spiral galaxies. We took the radial HI profiles from the literature and considered them up to distances from the center to which the available information allows us to estimatethe volume densities of HI, H$_2$, and stars as well as the thicknesses of the gaseous and stellar disks by the method of Narayan and Jog (2002). A brief description of the method, references to the radial HI profiles used, and the calculated volume densities and thicknesses of the disks are given in Abramova and Zasov (2008, 2011). When solving this problem, we took into account the dark halo, which plays a prominent role at the periphery of some galaxies.

In this paper, we reconstructed the total hydrogen profiles for 22 galaxies with radial HI profiles known from observations: ten low-surfacebrightness (LSB) galaxies, one normal-brightness galaxy with an extended LSB disk (NGC 289), three S0 galaxies, and seven normal galaxies (including our Galaxy). For comparison, we performed calculations for the spiral galaxy NGC 3198, which was considered previously by Maloney (1993) and Maloney and Bland-Hawthorn (1999). For this galaxy, we used the radial column density profiles and the hydrogen disk half-thickness from Maloney (1993). The list of the galaxies and the adopted distances to them are given in the Table~\ref{tab1}.
\begin{table}[tbp]
\setcaptionmargin{0mm} \onelinecaptionsfalse
\captionstyle{flushleft}\caption[Information about the sample galaxies]{
\label{tab1} Information about the sample galaxies}
\begin{center}
\bigskip
\begin{tabular}{c|c|c}
\hline
Galaxy&Type&$D$,~Mpc\\
\hline
~(1)~&~(2)~&~(3)~\\
\hline
Malin 1&LSB&330\,(Moore and Parker 2007)\\
Malin 2\,(F568-6)&LSB&186\,(Pickering et al. 1997)\\
F561-1&LSB&62.7\,(de Block et al. 1996)\\
F574-1&LSB&96\,(Swaters et al. 2003)\\
F568-1&LSB&85\,(Swaters et al. 2003)\\
F568-v1&LSB&80\,(Swaters et al. 2003)\\
F568-3&LSB&77\,(de Block et al. 2001)\\
UGC 128&LSB&60\,(van der Hulst et al. 1993)\\
UGC 1230&LSB&51\,(de Block and Bosma 2002)\\
UGC 6614&LSB&85\,(de Block et al. 2001)\\
NGC 289&~~SBbc+LSB~~&21\,(Walsh et al. 1997)\\
UGC 2487\,(NGC 1167)&S0&67.4\,(Noordermeer 2006)\\
UGC 11670\,(NGC 7013)&S0&12.7\,(Noordermeer 2006)\\
UGC 11914\,(NGC 7217)&S0&14.9\,(Noordermeer 2006)\\
M33\,(NGC 598)&~~Sc~~&0.7\,(Boissier et al. 2004)\\
M51\,(NGC 5194)&Sbc&8.4\,(Boissier et al. 2004)\\
M81\,(NGC 3031)&Sab&3.6\,(Boissier et al. 2004)\\
M100\,(NGC 4321)&SABb&17.0\,(Boissier et al. 2004)\\
M101\,(NGC 5457)&SABc&7.5\,(Boissier et al. 2004)\\
M106\,(NGC 4258)&SABb&8.0\,(Boissier et al. 2004)\\
Galaxy&SBbc&~~---~~\\
NGC 3198&SB(rs)c&9.4\,(Maloney 1993)\\
\hline
\end{tabular}
\end{center}
\begin{flushleft}
\end{flushleft}
\end{table}

Here, we consider the basic equations solved in the problem, present the results of our calculations, and calculate the radial profiles of atomic and ionized gas for a given, monotonically decreasing total gas profile. In final section “Discussion and Conclusions” we analyze the obtained results.

\section{BASIC EQUATIONS}

\hspace{0.6cm}The main adopted simplifications are as follows. The HI disk is assumed to be an equilibrium and axisymmetric one, and the thickness of the stellar disk is determined by the condition of gravitational stability that imposes constraints on the stellar velocity dispersion. In addition, we disregard the local gas density inhomogeneities, whose presence changes the gas ionization fraction.

The ionization balance equation for the gas layer at distance $z$ from the disk midplane ($z = 0$) is
\begin{equation}\label{1}
\alpha_B n_e n_p = n_\textrm{HI}
\int_{\nu_0}^\infty\frac{4\pi\,J_\nu(z)}{h\nu}\,\sigma_\nu\,d\nu,
\end{equation}
where $z$ is the coordinate along the axis perpendicular to the disk plane; $h\nu_0=13.6$~eV; $n_e$ and $n_p$ [cm$^{-3}$] are the ion and proton number densities, respectively; $n_\textrm{HI}$ [cm$^{-3}$] is the number density of hydrogen atoms; $\sigma_\nu=7\cdot10^{-18}\left(\frac{13.58}{h\nu}\right)^3$~[cm$^2$] is the photoionization cross section for the hydrogen atom; $J_\nu(z)~\left[\frac{\textrm{erg}}{\textrm{cm}^2\,\textrm{s\,Hz\,sr}}\right]$ is the flux of ionizing radiation at given $z$; $\alpha_B$ is the radiative recombination coefficient, which generally depends on the amount of neutral hydrogen and the local electron temperature. Since the contribution of elements heavier than hydrogen is small in our problem, below we assume that $n_e\approx n_p$.

According to Ferland (1980), the total recombination coefficient
\begin{equation*}
\alpha_B = \sum_{n=2}^\infty \alpha_n
\end{equation*}
can be approximated to within 10\% by the expression
\begin{equation*}
\alpha_B(T_e) = \left\{ \begin{array}{ll}
2.9\cdot10^{-10}~T_e^{-0.77},~&T_e\leq 2.6\cdot 10^4\,\textrm{K};\\
1.31\cdot10^{-8}~T_e^{-1.13},~&T_e> 2.6\cdot 10^4\,\textrm{K}.
\end{array} \right.
\end{equation*}
In our case, $T_e=10^4$~K, so that $\alpha_B=2.4\cdot10^{-13}$
cm$^3$s$^{-1}$.

The initial flux of ionizing radiation $J_\nu^0$ (uncorrected for absorption) can be written as (see Bland-Hawthorn et al. 1997)
\begin{equation}\label{2}
J_\nu^0=10^{-21}\,J_{-21}^0\,\left(\frac{\nu_0}{\nu}\right)^\beta~\left[\frac{\textrm{erg}}{\textrm{cm}^2\,\textrm{s\,Hz\,sr}}\right],
\end{equation}
where $J_{-21}^0=0.08~\frac{\textrm{erg}}{\textrm{cm}^2\,\textrm{s\,Hz\,sr}}$ is the metagalactic flux near the Lyman limit $\nu=\nu_0$. If the cosmic radiation is determined by quasars at high redshifts, then, according to Bland-Hawthorn et al. (1997), $\beta\approx 1$, i.e., $J_\nu^0=0.08\cdot10^{-21}\left(\frac{\nu_0}{\nu}\right)~\frac{\textrm{erg}}{\textrm{cm}^2\,\textrm{s\,Hz\,sr}}$.

When penetrating into the disk, part of the initial flux $J_\nu^0$ is absorbed and the attenuated flux at distance $z$ from the disk midplane will be
\begin{equation}\label{3}
J_\nu(z)=J_\nu^0\,e^{-\sigma_\nu\int_z^\infty n_{\textrm{HI}} (y)\,dy}+J_\nu^0\,e^{-\sigma_\nu\int_{-\infty}^z n_{\textrm{HI}}(z')\,dz'}.
\end{equation}

We will assume that the number densities $n_e=n_p$ and $n_\textrm{HI}$ decrease along the $z$ coordinate exponentially:
\begin{equation}\label{4}
n_{p,\textrm{HI}}=n_{p,\textrm{HI}}^0\,e^{-\frac{z}{z_s}},
\end{equation}
where the vertical scale height of the disk $z_s$ is related to the half-thicknesses of the disks of the sample galaxies $h_{1/2}$ [pc] by the relation
\begin{equation}\label{5}
z_s=3.086\cdot10^{18}\,\frac{h_{1/2}}{\ln 2}.
\end{equation}
The values of $h_{1/2}(z)$ for the galaxies considered here were calculated previously (Abramova and Zasov 2008, 2011).

For the convenience of our calculations, we will change the system of units and pass from the particle number densities ni to the column densities $N_i$ [cm$^{-2}$], which are related to the observed quantities, the surface density $\sigma_i$, by the relation
\begin{equation}\label{51}
N_i=1.26\times10^{20}\sigma_i,~{\textrm{where}}
\end{equation}
$\sigma_i$ are measured in ${\cal M}_\odot$/pc$^2$. Let us express $N_i$ in terms of the particle number densities $n_i$ (the subscript $i$ can refer to both protons $p$ and neutral HI atoms):
\begin{equation}\label{58}
N_i=\int_{-\infty}^{\infty}\,n_i(z)\,dz=2\int_0^\infty \,n_i(z)\,dz=2n_i^0\int_0^\infty
e^{-\frac{z}{z_s}}\,dz=2n_i^0z_s,
\end{equation}
where $n_i^0$ are the number densities of the corresponding particles in the disk midplane at $z=0$. Thus, if $n_p$ and $n_\textrm{HI}$ change exponentially along the $z$ coordinate, then $N_i$ is expressed linearly in terms of $n_i$. Since the ionization balance equation~(\ref{1}) is valid for all $z$, we will consider it for the case of $z=0$.

\subsection{Equation~(\ref{1}) for the Disk Midplane}

\hspace{0.6cm}For $z=0$, the flux of ionizing radiation~(\ref{3}) attenuated by absorption can be written as
\begin{equation}\label{6}
J_\nu(z)=J_\nu(0)=2J_\nu^0e^{-\sigma_\nu n_{\textrm{HI}}^0z_s}.
\end{equation}
The factor 2 on the right-hand side of~(\ref{6}) means that the flux comes from two directions. In the ionization balance equation~(\ref{1}), we will pass from integration over the frequency to integration over the energy $E=h\nu$ and from the particle number densities $n_i$ to the proton (ionized gas) and HI surface densities. Given (\ref{51}), (\ref{5}), and (\ref{6}), we will write (\ref{1}) for $\sigma_p$:
\begin{equation}\label{7}
\sigma_p^2=2.1\cdot10^4\sigma_{\textrm{HI}}\,h_{1/2}\,\int_{13.6}^\infty\frac{e^{-1.1\cdot10^6\,\frac{\sigma_{\textrm{HI}}}{E^3}}}{E^5}\,dE.
\end{equation}

Let us denote $A=1.1\cdot10^6\,\sigma_{\textrm{HI}}$ and $E_0=13.6$~eV. The integral on the right-hand side of~(\ref{7}) will then be rewritten as
\begin{equation}\label{8}
\int_{E_0}^\infty \frac{e^{-\frac{A}{E^3}}}{E^5}\,dE=
\frac{1}{3A^{4/3}}\,\int_{0}^{\frac{A}{E_0^3}}\sqrt[3]{z'}\,e^{-z'}\,dz'.
\end{equation}
According to Gradshtein and Ryzhik (1963),
\begin{equation*}
\int_0^u x^{\nu-1}\,e^{-\mu x}\,dx=\mu^{-\nu}\,\gamma(\nu;\mu u),
\end{equation*}
where $\gamma(\nu;\mu u)$ is the incomplete gamma function.

The final ionization balance equation in the disk midplane written for the sought-for quantity, the proton (ionized gas) surface density, will take the form
\begin{equation}\label{9}
\sigma_p^2=7\cdot10^3~\frac{\sigma_{\textrm{HI}}\,h_{1/2}}{\sqrt[3]{A^4}}~\gamma\left(\frac43;\frac{A}{E_0^3}\right),~A=1.1\cdot10^6\,\sigma_{\textrm{HI}}.
\end{equation}
The right-hand side of~(\ref{9}) depends only on two quantities: observed surface density of atomic hydrogen $\sigma_{\textrm{HI}}$ and the half-thicknesses of the gaseous disks of the galaxies considered $h_{1/2}$, which were found for all galaxies of the sample previously (Abramova and Zasov 2008, 2011).

Equation~(\ref{9}) is solved by numerical methods.

\section{THE RADIAL PROFILES OF TOTAL (NEUTRAL + IONIZED) AND IONIZED HYDROGEN}

\begin{figure}
\centering
{\includegraphics[scale=0.50]{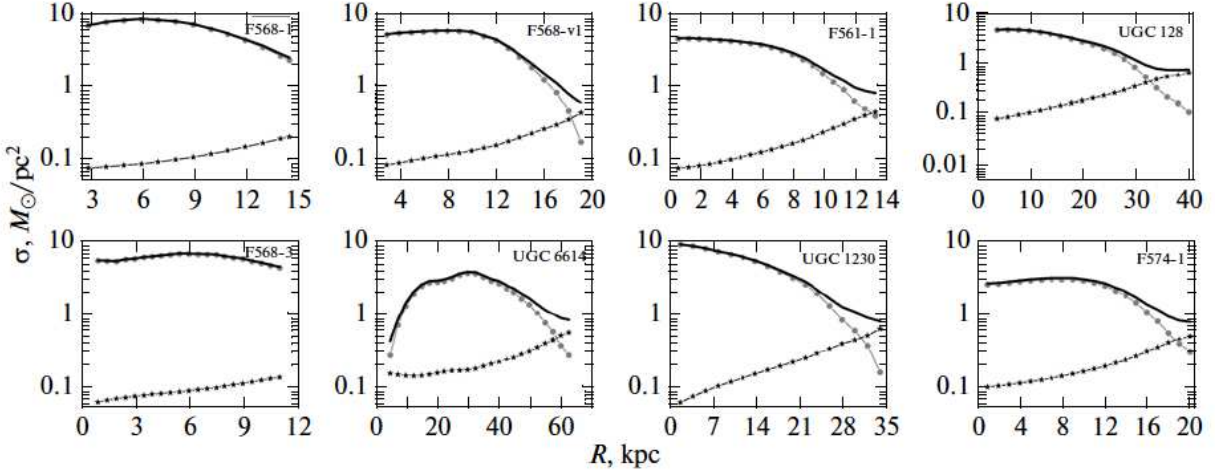}}
\caption{\footnotesize Modeling of the radial surface density distribution for ionized and total (ionized+HI) gas based on the observed radial HI surface density distribution for LSB galaxies: the solid line, black asterisks, and gray circles indicate, respectively, the total gas density, the ionized gas distribution, and the radial HI distribution.}\label{fig1}
\end{figure}

\begin{figure}
\centering
{\includegraphics[scale=0.49]{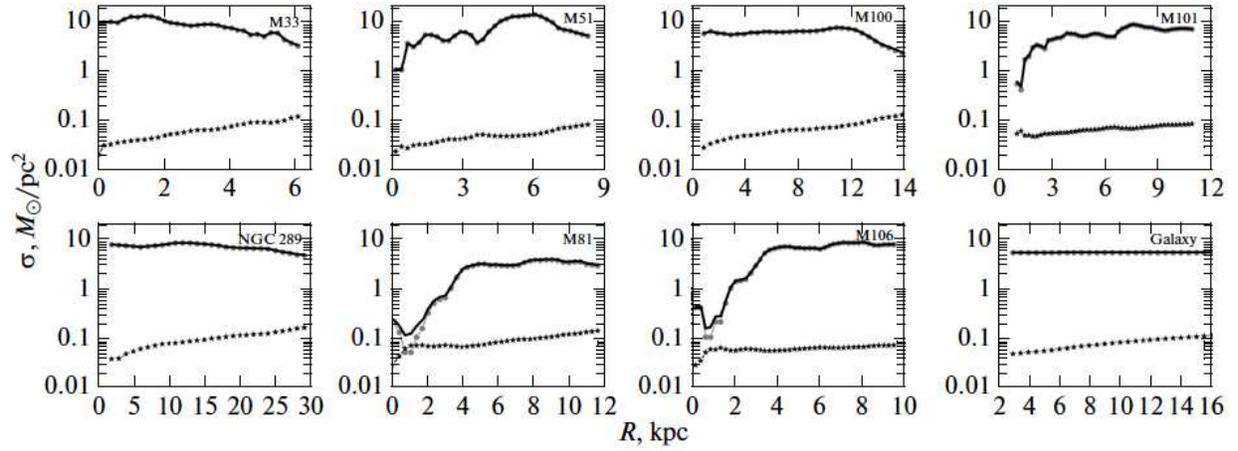}}
\caption{\footnotesize Same as Fig.~(\ref{fig1}) for normal galaxies and the normal+LSB galaxy NGC~289.}\label{fig2}
\end{figure}

\begin{figure}
\centering
{\includegraphics[scale=0.50]{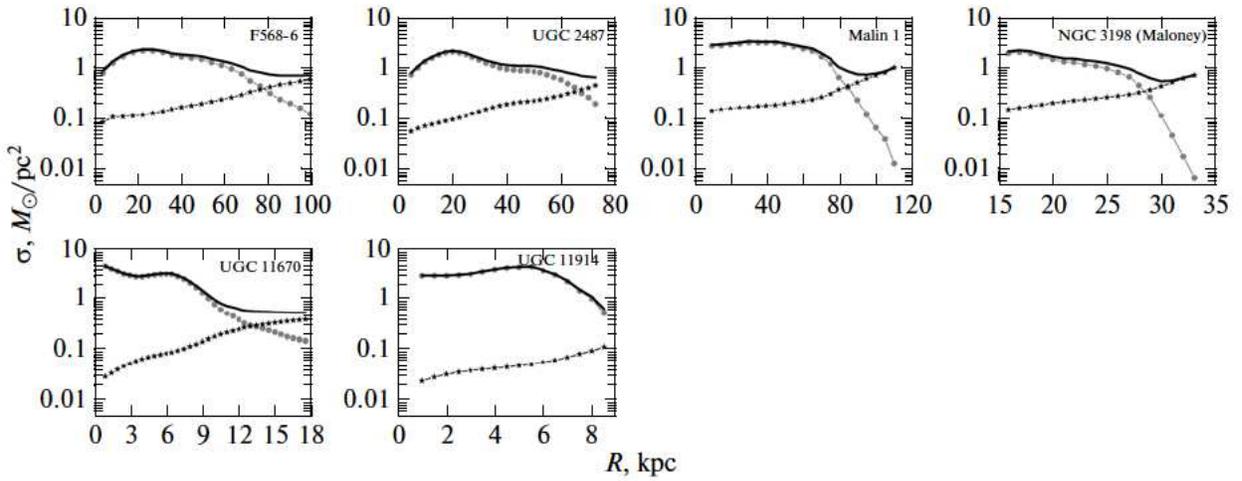}}
\caption{\footnotesize Same as Figs.~(\ref{fig1}) and (\ref{fig2}) for S0 galaxies (UGC 2487, UGC 11670 and UGC 11914), two giant LSB galaxies, Malin 1 and F568-6 (Malin 2), and the spiral galaxy NGC~3198 considered previously by Maloney (1993) and Maloney and Bland- Hawthorn (1999).}\label{fig3}
\end{figure}

\hspace{0.6cm}Figures~(\ref{fig1}-\ref{fig3}) present the modeled radial surface density profiles for ionized and total (neutral + ionized) gas. The objects of all types belong to the galaxies that have a cutoff in the observed radial HI profile: LSB (F561-1, F568-v1, F568-6, F574-1, UGC 128, UGC 1230, UGC 6614, Malin 1), S0 (UGC 2487, UGC 11914), and normal galaxies (M33, M100, and NGC 3198). For M81 there may be a cutoff in the radial HI profile at the boundary of the region considered here, while for the remaining galaxies a gradual decrease in the amount of atomic hydrogen with increasing distance from the center is predominantly observed.

We see from Figs.~(\ref{fig1}-\ref{fig3}) that the amounts of ionized and neutral gas for the galaxies of all types become equal in the range of surface densities $\sigma_{\textrm{HI}}=0.3\div 0.5~{\cal M}_\odot/{\textrm pc}^2$ ($\sigma_{\textrm{tot}}=0.6\div 1.0~{\cal M}_\odot/{\textrm pc}^2$), i.e., when the column density reaches $N_{\textrm{HI}}=3.8\cdot 10^{19}\div 6.3\cdot 10^{19}~{\textrm cm}^{-2}$. Both among the models proposed by Maloney (1993) for NGC~3198 and among the models computed by Corbelli and Salpeter (1993) for M33, there are those that are consistent with this result.

For the normal galaxies (Fig.~\ref{fig2}) in the $R$ range considered here, the relative ionized gas fraction is small. According to Narayan and Jog (2002), the amount of HI in the range $R=3-16$~kpc in our Galaxy (Fig.~\ref{fig2}) is almost constant and the ionization grows weakly. The behavior of the giant normal+LSB galaxy NGC~289 is identical to that of the normal galaxies considered. The S0 galaxies UGC~2487, UGC~11670, and UGC~11914 (Fig.~\ref{fig3}), as in the case of densities (see Abramova and Zasov 2011), are closer in their properties to the LSB galaxies.

In the case of LSB and S0 galaxies, for which the gas ionization is more significant (see Figs.~(\ref{fig1}) and (\ref{fig3})), and in the normal galaxy M100 (see Fig.~\ref{fig2}), the ionization fraction near the cutoff in the observed radial HI profile is low. This may suggest that the sharp drop in the observed HI density is caused not by ionization but by a sharp drop in the initial gas density.

\section{CALCULATION OF THE RADIAL HI AND IONIZED GAS PROFILES FROM THE KNOWN TOTAL GAS DENSITY PROFILE}

\subsection{The Model Galaxy}

\hspace{0.6cm}Suppose that the initial density profile is known. Let us write the ionized gas surface density as $\sigma_p=\sigma_{\textrm{tot}}-\sigma_\textrm{HI}$. We will assume the total gas density
($\sigma_p+\sigma_\textrm{HI}$) to be known. We solve~(\ref{9}) numerically for $\sigma_\textrm{HI}$ and use the value found to calculate the radial ionized gas surface density profile.

\begin{figure}[h]
\centering
\includegraphics[scale=0.50]{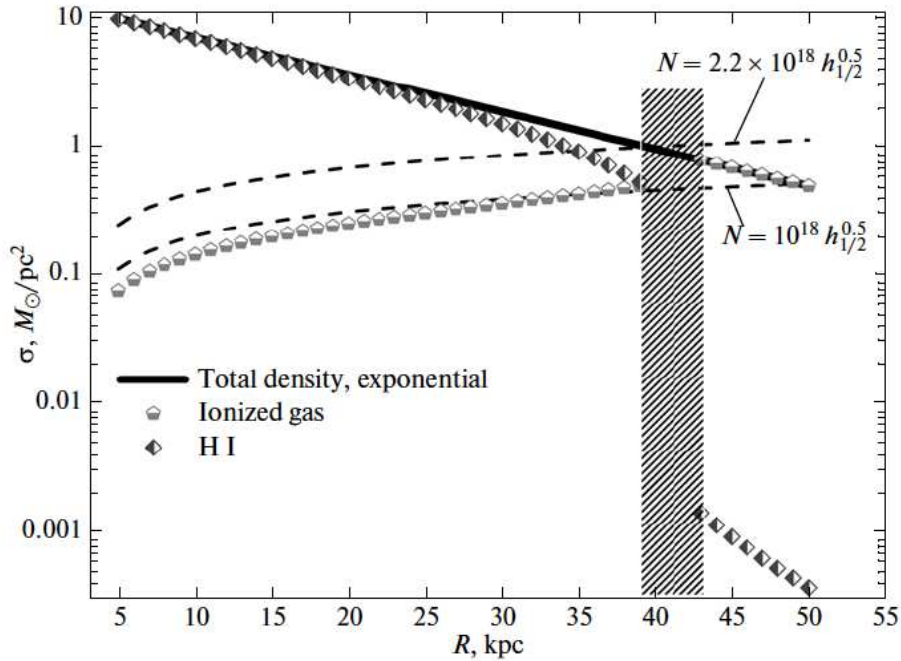}
\caption{\footnotesize Modeling of the radial atomic and ionized hydrogen profiles for an exponential initial gas surface density distribution. The dashed lines indicate the critical densities; the hatching marks the regions of mathematical instability of the solution (see the text).}\label{fig4}
\end{figure}

Consider a model galaxy whose parameters are chosen arbitrarily based on typical values for the normal galaxies considered here. Let the total gas surface density in the outer region of the model galaxy drop exponentially
\begin{equation}\label{10}
\sigma_{\textrm{tot}},~[{\cal M}_\odot/\textrm{pc}^2]=10e^\frac{5-R}{15},~R>5~\textrm{kpc},
\end{equation}
and the half-thickness of the gaseous disk increase linearly with radius
\begin{equation}\label{11}
h_{1/2}(R),~[\textrm{pc}]=90R-250.
\end{equation}
Figure~(\ref{fig4}) presents the results of our calculations. The dashed lines indicate the column densities at which (according to the estimates from Abramova and Zasov (2011)) hydrogen turns out to be ionized over the entire thickness of the gas layer (the socalled critical density). We see that in the range of critical densities the mass fraction of ionized hydrogen is 30–50\%.

The cutoff in the theoretical HI profile begins where the densities of atomic and ionized hydrogen in the model galaxy become equal. This occurs near $\sigma_{\textrm{HI}}\approx 0.5~{\cal M}_\odot/{\textrm pc}^2$, which is in good agreement with the results obtained for the 22 galaxies considered here.
When the amounts of atomic and ionized hydrogen become equal, a transition zone with an extent of several kpc in which the solution is mathematically unstable begins (hatched in Fig.~(\ref{fig4}). In the transition zone, the gas ionization fraction increases sharply with distance from the center (see Fig.~(\ref{fig4})).

\subsection{Examples of Modeling Several Galaxies}

\hspace{0.6cm}As an illustration, let us solve the inverse problem considered above for several galaxies with farreaching $\sigma_\textrm{HI}$ profiles. Let us take the two LSB galaxies UGC~128 and UGC~1230, the giant lenticular galaxy UGC~2487, and the spiral galaxy NGC~3198. We will specify the radial total gas profiles in the form of an exponential with a radial scale length corresponding to the observed profile in the inner region of the galaxy and will solve Eq.~(\ref{9}) for HI.

\begin{figure}[h]
\centering
\includegraphics[scale=0.50]{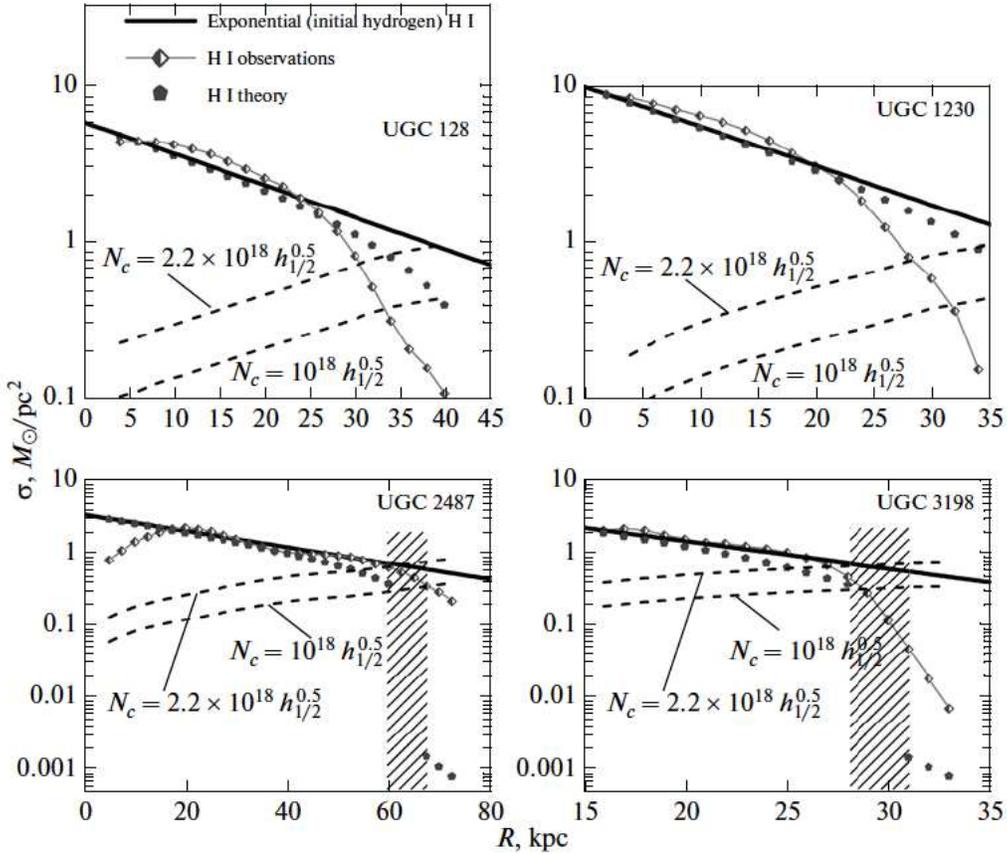}
\caption{\footnotesize Reconstruction of the radial HI profile from the specified exponential total gas profile for the four galaxies. The dashed lines indicate the critical densities; the regions of mathematical instability of the solution are shaded (see the text).}\label{fig5}
\end{figure}

For each of the four galaxies, Fig.~(\ref{fig5}) shows the model exponential initial gas profiles as well as the observed and theoretical atomic hydrogen profiles.

We see from Fig.~(\ref{fig5}) that for two galaxies (the giant S0 UGC~2487 and the spiral NGC~3198), when the densities reach their critical values N$_\textrm{C}$, the HI densities change abruptly, just as in the model galaxy considered in the previous section. This occurs near
$\sigma_{\textrm{HI}}\approx 0.4~{\cal M}_\odot/{\textrm pc}^2$, which is also in good agreement with the results obtained for the 22 galaxies considered here and with the results obtained for the model galaxy.

For three galaxies (UGC~128, UGC~1230, and UGC~2487), the HI cutoff profiles do not correspond to that expected under ionization: for the LSB galaxies UGC~128 and UGC~1230, which are characterized by higher HI surface densities, the model curves of the radial HI distribution lie on the graphs above
the observed ones and, in contrast to the observed curves, have no distinct cutoff, while the theoretical HI profile for the giant S0 UGC~2487 lies below the observed one and, in contrast to it, has a jump. The result of modeling NGC~3198 agrees well with the observations.

Thus, in the galaxies where the HI surface densities at the edge of the optical disk reach $\sigma_{\textrm{HI}}\le 0.5~{\cal M}_\odot/{\textrm pc}^2$, the sharp cutoff in the observed HI profile can be caused by the ionization of gas by background radiation. At the same time, for galaxies with higher surface densities at the disk edge, the steepening of the observed HI profile begins closer to the center than could be expected in the case of ionization by background radiation and, therefore, must be caused by other factors. Note that Portas et al. (2009), who analyzed the distribution of atomic gas in nine galaxies, made a similar suggestion: the radial HI profiles are probably determined by internal causes rather than external ones, to which the background radiation belongs.

\section{DISCUSSION AND CONCLUSIONS}

\hspace{0.6cm}When solving the problem, we made a number of simplifications. We assumed that there were no inhomogeneities in the gaseous disk (allowance for them reduces the ionization fraction). In addition, we disregarded the fact that for the regions deep in the disk, the ionizing radiation loses its isotropy. These simplifications could lead the effect that is observed in the galaxies Malin~1 and NGC~3198 (Fig.~\ref{fig3}), for which the total gas profile at the disk edge turns out to be bent upward. This bend is unlikely to be real; this is most likely the result of an overestimation of
the ionized gas mass at the edges of the disks of these galaxies.

The overestimation of the ionized gas density can be not only the result of the above simplifications
but also, in the case of Malin~1, a consequence of the peculiarities of the numerical method used to
determine the gaseous disk thickness (see Abramova and Zasov 2011). Indeed, in the regions of very low densities $<10^{-27}~{\textrm g}/{\textrm {cm}^3}$ characteristic of the edge of the Malin~1 disk, the accuracy of the numerical method used is insufficient. This could lead to an overestimation of $h_{1/2}$ at the disk edge and, as a consequence, to an overestimation of the ionization and a bend of the theoretical total gas density profile. The possibility that the half-thickness of the gaseous
disk of NGC~3198 taken from Maloney (1993) was overestimated must not be ruled out either. This could also be responsible for bend of the theoretical total gas density profile in this galaxy.

The results of solving the direct and inverse problems for galaxies of different types (LSB, S0, and normal) lead us to the following conclusion: for galaxies in which the sharp steepening of the radial HI profile begins in regions with densities $\sigma_{\textrm{HI}}\le 0.5~{\cal M}_\odot/{\textrm pc}^2$, it can actually be caused by the ionization of gas at the disk edge. In this paper, two of the galaxies considered here, the giant S0 UGC~2487 and the spiral NGC~3198, belong to such objects. At the same time, for the remaining galaxies with a cutoff in the radial HI profile, the sharp decrease in the observed amount of atomic hydrogen in the disk begins well before the gas surface density reaches $\sigma_{\textrm{HI}}\approx0.5~{\cal M}_\odot/{\textrm pc}^2$ and, consequently, the cutoff in the radial HI profile in them is not related to the presence of ionizing radiation but is attributable to other factors.

\begin{figure}[h]
\centering
\includegraphics[scale=0.45]{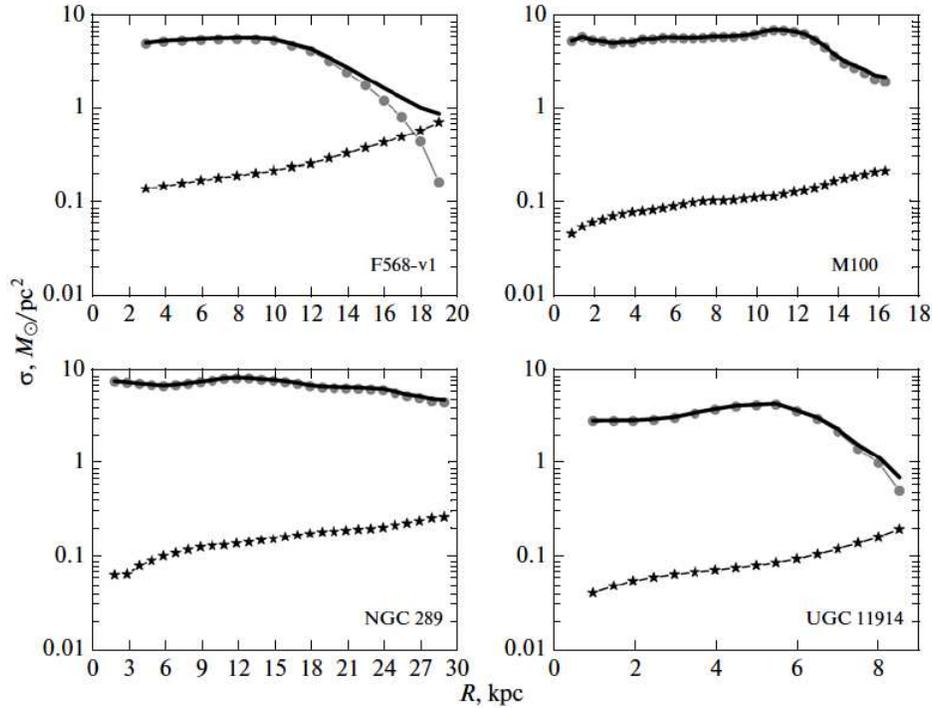}
\caption{\footnotesize Results of modeling the radial surface density distribution for ionized and total (ionized+HI) gas based on the observed radial HI surface density distribution for a threefold increase in the flux of ionizing radiation~(\ref{6}). The notation is the same as that in Figs.~(\ref{fig1}-\ref{fig3}).}\label{fig6}
\end{figure}

Even a threefold increase in the flux of ionizing radiation does not change the situation (see Fig.~\ref{fig6}).
This figure presents the solution of the direct problem for four galaxies of different types. The increase in the radiation flux had no noticeable effect on the ionization of gas in the normal galaxy M100, the normal+LSB galaxy NGC~289, and the S0 galaxy UGC~11914. At the same time, the ionized gas fraction in the LSB galaxy F568-v1 increased to an extent that its amount at the disk edge exceeded the amount of HI (cf. Fig.~\ref{fig1}). However, despite this fact, the result qualitatively remained the same for all four galaxies: the steepening of the observed HI profile begins closer to the center than might be expected in the case of background ionization.

The inhomogeneity of the interstellar medium is an important factor that can change the estimate of the gas ionization fraction in disk regions far from the center. The concentration of gas into clouds or filaments will reduce its ionization fraction, moving
the HI profile cutoff even farther from the center, and cannot explain the sharp decrease in density that begins earlier than that expected for themean density. However, this effect is difficult to estimate quantitatively because of the lack of sufficient information about the small-scale distribution of gas.

Basically, the increase in ionization fraction could be explained by internal sources of ionizing radiation (young stars, supernovae), but the available data for the galaxies under consideration give no grounds for this. Indeed, most of the galaxies with an observed steepening of the HI profile belong to lenticular or LSB galaxies, and, consequently, they have neither contrasting spiral arms nor intense star formation. Only three of these galaxies (M33, M100, and NGC~3198) belong to late morphological types; for the latter, no internal sources of ionization are required to explain the cutoff in the HI profile. In addition, the spiral structure at the periphery of galaxies (in the regions where the cutoffs are observed) is rather indistinct.

Note also that among the 22 galaxies considered, there are no objects located in clusters (except for
M100 at the periphery of the Virgo cluster) and only four galaxies (M81, M33, NGC 289, and UGC 2487) are members of groups. No cutoff in the HI distribution is observed in the range of radii considered in the galaxy with a close companion (M51).

Therefore, we cannot consider the ionization of a rarefied gas at the edges of galactic disks by external radiation as a universal mechanism leading to the observed cutoff in the radial HI profiles, and other explanations for this phenomenon related to the formation and evolution of disks should be sought.

\section{ACKNOWLEDGMENTS}
\hspace{0.6cm}I wish to thank A.V. Zasov and N.G. Bochkarev for fruitful discussions and valuable remarks made during my work on this paper and the anonymous referee whose comments helped improve the final section of the paper “Discussion and Conclusions”. This work was supported by the Russian Foundation for Basic Research (project no. 11-02-12247-ofi-m-2011).

\bf{Translated by G. Rudnitskii}

\end{document}